\documentclass[sigconf]{acmart}
\usepackage{amssymb,amsfonts}
\usepackage{graphicx}
\usepackage{algorithm}
\usepackage{algpseudocode} 
\usepackage{multirow}
\usepackage{subcaption}
\usepackage{enumitem}
\usepackage{textcomp} 
\usepackage{siunitx}   
\usepackage{soul}
\usepackage{xcolor}
\usepackage{url}

\newcommand{\etal}{{et al. }}
\newcommand{\eg}{{\it{e.g., }}}
\newcommand{\ie}{{\it{i.e., }}}

\newcommand{\tool}[0]{Uno}
\newcommand{\toolCC}[0]{\tool CC}
\newcommand{\toolLB}[0]{\tool RC}

\newcommand{\encircle}[1]{%
  \scalebox{1.4}{\textcircled{\scriptsize #1}}%
}

\definecolor{mycol}{RGB}{214,0,147}

\AtBeginDocument{%
  }

\copyrightyear{2025}
\acmYear{2025}
\acmYear{2025}
\setcopyright{cc}
\setcctype{by}
\acmConference[SC '25]{The International Conference for High Performance
Computing, Networking, Storage and Analysis}{November 16--21, 2025}{St
Louis, MO, USA}
\acmBooktitle{The International Conference for High Performance Computing,
Networking, Storage and Analysis (SC '25), November 16--21, 2025, St Louis,
MO, USA}
\acmDOI{10.1145/3712285.3759884}
\acmISBN{979-8-4007-1466-5/2025/11}

\begin{document}

\title{Uno: A One-Stop Solution for Inter- and Intra-Datacenter Congestion Control and Reliable Connectivity}

\author{Tommaso Bonato}
\authornote{Both authors contributed equally to this research.}
\affiliation{%
  \institution{ETH Z\"urich}
  \city{Z\"urich}
  \country{Switzerland}
}
\affiliation{%
  \institution{Microsoft}
  \city{Redmond}
  \country{USA}
}
\email{tommaso.bonato@inf.ethz.ch}

\author{Sepehr Abdous}
\authornotemark[1]
\affiliation{%
  \institution{Johns Hopkins University}
  \city{Baltimore}
  \country{USA}
}
\affiliation{%
  \institution{Microsoft}
  \city{Redmond}
  \country{USA}
}
\email{sabdous1@jh.edu}

\author{Abdul Kabbani}
\affiliation{%
  \institution{Microsoft}
  \city{Redmond}
  \country{USA}
}
\email{abdulkabbani@microsoft.com}

\author{Ahmad Ghalayini}
\affiliation{%
  \institution{Microsoft}
  \city{Redmond}
  \country{USA}
}
\email{aghalayini@microsoft.com}

\author{Nadeen Gebara}
\affiliation{%
  \institution{Microsoft}
  \city{Redmond}
  \country{USA}
}
\email{nadeengebara@microsoft.com}

\author{Terry Lam}
\affiliation{%
  \institution{Microsoft}
  \city{Redmond}
  \country{USA}
}
\email{thelam@microsoft.com}

\author{Anup Agarwal}
\affiliation{%
  \institution{Carnegie Mellon University}
  \city{Pittsburgh}
  \country{USA}
}
\email{anupa@cmu.edu}

\author{Tiancheng Chen}
\affiliation{%
  \institution{ETH Z\"urich}
  \city{Z\"urich}
  \country{Switzerland}
}
\email{tiancheng.chen@inf.ethz.ch}

\author{Zhuolong Yu}
\affiliation{%
  \institution{Microsoft}
  \city{Redmond}
  \country{USA}
}
\email{zhuolongyu@microsoft.com}

\author{Konstantin Taranov}
\affiliation{%
  \institution{Microsoft}
  \city{Redmond}
  \country{USA}
}
\email{kotaranov@microsoft.com}

\author{Mahmoud Elhaddad}
\affiliation{%
  \institution{Microsoft}
  \city{Redmond}
  \country{USA}
}
\email{maelhadd@microsoft.com}

\author{Daniele De Sensi}
\affiliation{%
  \institution{Sapienza University}
  \city{Rome}
  \country{Italy}
}
\email{desensi@di.uniroma1.it}

\author{Soudeh Ghorbani}
\affiliation{%
  \institution{Johns Hopkins University}
  \city{Baltimore}
  \country{USA}
}
\email{soudeh@soudeh.net}

\author{Torsten Hoefler}
\affiliation{%
  \institution{ETH Z\"urich}
  \city{Z\"urich}
  \country{Switzerland}
}
\email{htor@inf.ethz.ch}

\renewcommand{\shortauthors}{Bonato, Abdous, \etal}

\begin{abstract}

Cloud computing and AI workloads are driving unprecedented demand for efficient communication within and across datacenters. However, the coexistence of intra- and inter-datacenter traffic within datacenters plus the disparity between the RTTs of intra- and inter-datacenter networks complicates congestion management and traffic routing. Particularly, faster congestion responses of intra-datacenter traffic causes rate unfairness when competing with slower inter-datacenter flows. Additionally, inter-datacenter messages suffer from slow loss recovery and, thus, require reliability. Existing solutions overlook these challenges and handle inter- and intra-datacenter congestion with separate control loops or at different granularities. We propose \tool, a unified system for both inter- and intra-DC environments that integrates a transport protocol for rapid congestion reaction and fair rate control with a load balancing scheme that combines erasure coding and adaptive routing. Our findings show that \tool\ significantly improves the completion times of both inter- and intra-DC flows compared to state-of-the-art methods such as Gemini.
\end{abstract}


\settopmatter{printacmref=false} 
\pagestyle{plain}                
\fancyhead{}                     
\acmConference[]{}{}{}
\acmBooktitle{}
\acmISBN{}
\acmDOI{}

\maketitle

\section{Introduction}
\label{sec:intro}
With the drastic growth in cloud computing, HPC, and AI workloads, ensuring congestion-free communication and efficient traffic routing both inside and across multiple datacenters (DCs) is becoming more crucial than ever \cite{wan_importance,wan_importance2,tina}. Specifically, reports from Google highlight a $100\times$ increase in their inter-DC WAN traffic volume during a five-year period \cite{google_wan}.
Additionally, with the growth of large-scale AI models, fitting entire training jobs inside a single datacenter is becoming infeasible \cite{luo2024efficient}, \eg Google's Gemini was trained on several Google supercomputers \cite{gemini_google}, and, more recently, OpenAI used several clusters to train its GPT-4.5 model \cite{openai_yt}.

Many congestion control protocols have been developed throughout the years to separately ensure efficient communication in intra-DC \cite{dctcp, swift, bolt, d2tcp, mprdma, hpcc, timely, poseidon, powerTCP} and inter-DC \cite{bbr, swan, wan_importance, onewan, anulus} environments but very little research has been done on simultaneously handling both inter-DC and intra-DC traffic \cite{gemini}. While inter- and intra-datacenter traffic are usually treated as separate entities, they co-exist within datacenters and compete over resources. Therefore, it is crucial to ensure efficient communication for each entity while also considering the other entity \cite{gemini}.
However, doing so is challenging due to the inherent differences between datacenter networks and WANs. In particular, within a single datacenter, cable lengths and propagation delays are small and, mostly, homogeneous. However, inter-datacenter WANs are built using long physical links with large propagation delays \cite{gemini, anulus}. Additionally, inter-DC links traverse different geographical paths thereby introducing heterogeneity in link propagation delays and additional risks of failures.

\begin{figure}[t]
    \centering
    \includegraphics[width=0.99\linewidth]{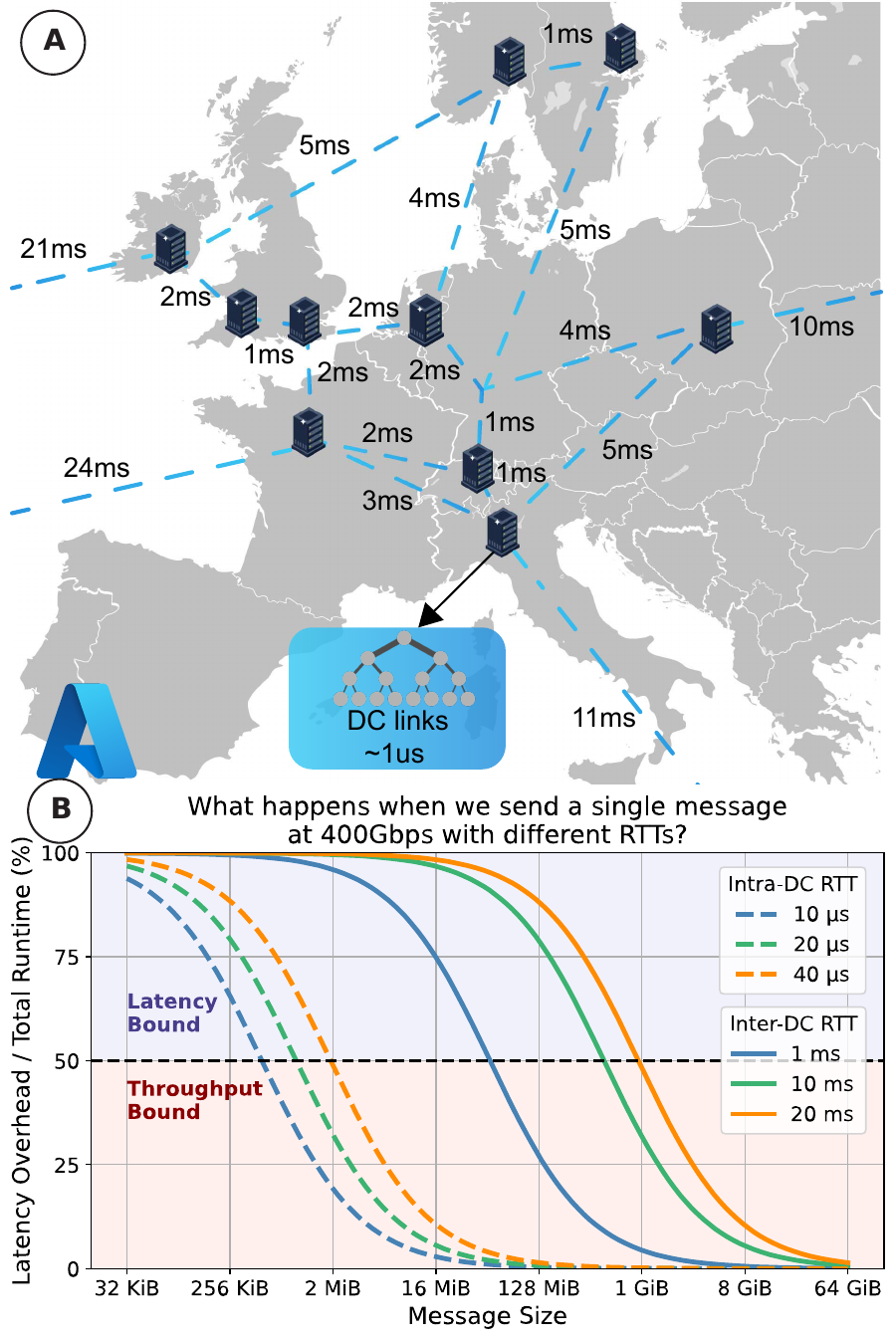}
    \vspace{-2mm}
    \caption{\small{\protect\encircle{A} shows inter-DC links for Azure in Europe and their delay assuming point-to-point connections. \protect\encircle{B} shows that inter-DC links make even medium-large messages latency-bound.}}
    \vspace{-6mm}
    \label{fig:poster}
\end{figure}


Unlike datacenter networks, in inter-DC WANs, the completion time of even large messages is bounded by latency rather than throughput due to the large propagation delays. Particularly, in a modern datacenter infrastructure, it might take at most tens of microseconds for a packet to go from any given node to any destination, assuming a lightly-loaded network \cite{clous_noise, timely}. On the other hand, when going across datacenters, such delay increases dramatically to multiple milliseconds, \eg Figure \ref{fig:poster} \encircle{A} presents link delays between Microsoft Azure's datacenters across Europe. Figure \ref{fig:poster} \encircle{B} presents the percentage of a message's completion time, \ie time taken from sending the first packet of the message to receiving the last ACK, that is due to the aggregate propagation delay across distinct message sizes and intra- and inter-DC propagation delays (indicated as RTTs).
For typical intra-datacenter RTTs (\ie $10\mu s$ to $40 \mu s$ \cite{swift, timely}), as we increase the message size, the completion time quickly becomes dominated by the sending throughput for sizes greater than 256 KiB. On the other hand, for inter-datacenter RTTs (\ie 1 ms to 60 ms \cite{gemini, anulus}), the completion time becomes mostly bounded by the propagation delay. For instance, when the inter-DC RTT is 20 ms, the completion time is dominated by propagation delay if messages are smaller than 1 GiB, which is quite large (message sizes recorded from Alibaba's inter-DC traces are all smaller than 300 MB \cite{flash_pass_alibaba}). Moreover, with increasing link bandwidths, this will only become more extreme in the future. While this theoretical study gives us some useful insights, we note that actual bounds would slightly change depending on the network's conditions.

This massive delay gap between intra-DC and inter-DC networks introduces several challenges for simultaneous and efficient congestion management of both intra- and inter-DC traffic: 

\textbf{1) Diverse congestion feedback granularity:} \textit{The congestion feedback loop for inter-DC flows is significantly delayed compared to intra-DC traffic.} Consequently, upon receiving a congestion signal for inter-DC traffic, it is hard to know if the path is still congested. On top of that, the delay mismatch makes it hard to maintain fairness between intra- and inter-DC flows while competing over a bottleneck link.

\textbf{2) BDP heterogeneity:} \textit{Inter-DC Bandwidth Delay Product (BDP) is significantly larger than intra-DC BDP due to its long RTTs}, \eg with 10 ms RTT and 400 Gbps link bandwidth, the inter-DC BDP is $\approx 500$ MB. While commodity switch buffers have increased in size during the years \cite{bai2017congestion}, they are still quite small, especially intra-DC switch buffers, compared to inter-DC BDP. Meanwhile, many congestion control algorithms \cite{dctcp, d2tcp, dcqcn} assume the capacity of switches to be at least a fraction of BDP, \eg 17\% for DCTCP \cite{dctcp}. While this assumption holds for intra-datacenter flows, whose BDPs are typically less than 1 MiB, it becomes unrealistic for inter-DC flows. The lack of buffering space in switches is amplified as cloud providers are using shallow-buffered commodity switches to be cost-efficient and improve scalability \cite{wan_importance, commodity1}. 

\textbf{3) Inefficient loss handling:} \textit{Packet loss and its consequent re-transmission significantly increase the message completion time in latency-bound WANs}. Even with advanced loss detection mechanisms such as packet trimming \cite{ndp}, the loss notification mechanism takes a long time due to the large propagation delay. Therefore, deploying efficient loss recovery mechanisms traffic is necessary.

\begin{figure}[t]
    \centering
    \includegraphics[width=0.9\linewidth]{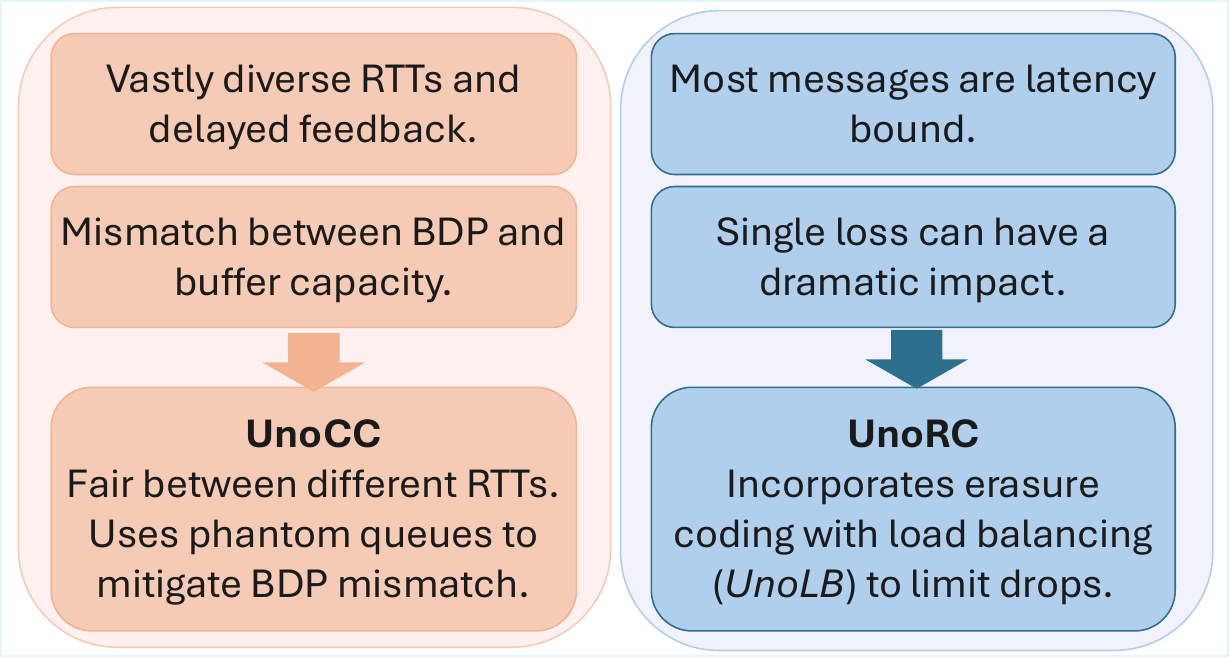}
    \vspace{-2mm}
    \caption{\small{To ensure efficient communication within and across datacenters, \tool\ integrates congestion control, load balancing, and loss resiliency.}}
    \vspace{-3mm}
    \label{fig:uno_intro}
\end{figure}

While existing proposals, such as Gemini \cite{gemini} and BBR \cite{bbr}, try to tackle some of these challenges, to the best of our knowledge, no proposals have addressed them all. Specifically, BBR is dedicated only to WAN traffic and requires another transport, such as DCTCP \cite{dctcp}, to handle intra-DC traffic. With separate transports, it is challenging to guarantee fairness. Gemini, on the other hand, is a window-based congestion control for both intra- and inter-DC communication that is proven to achieve bandwidth fairness among intra- and inter-DC flows. However, as Gemini's granularity for reacting to intra-DC and inter-DC congestion signals significantly varies, Gemini experiences slow convergence to fairness and is prone to network under-utilization (\S\ref{sec:motiv}). Lastly, all these techniques suffer from inefficient loss handling for WAN traffic. To resolve these problems, we introduce \tool, a system that tightly integrates congestion control, load balancing, and loss resiliency to create a unified solution for both intra- and inter-DC communication. As shown in Figure \ref{fig:uno_intro}, \tool\ employs two key components:

\textbf{1) Congestion Control component (\toolCC):} When messages become throughput-bound and congestion control becomes vital, \toolCC\ exploits Explicit Congestion Notification (ECN) to efficiently handle congestion both inside and across datacenters and simultaneously provide low latency and flow-level fairness for intra-DC and inter-DC traffic. To ensure effective congestion management across datacenters, \tool\ uses phantom queues \cite{phantomQ}, \ie virtual queues with arbitrary sizes that mimic the behavior of physical queues, to match the high BDPs of the inter-DC connections regardless of the physical queue capacity.

\textbf{2) Reliable Connectivity component (\toolLB):} Since most messages are latency-bound when traversing inter-DC WAN, \tool\ augments message transmission with \textit{erasure coding} \cite{erasure} to increase loss recovery without packet re-transmission for cross datacenter communication. Furthermore, \tool\ integrates the erasure coding logic with a sub-flow level load balancing (\tool LB) scheme that leverages multi-pathing of modern networks.

We evaluate \tool\ using htsim simulations \cite{ndp} and compare it against MPRDMA+BBR (\cite{mprdma}+\cite{bbr}) and Gemini \cite{gemini}. Our results show that \tool\ significantly improves latency, fairness, and loss resiliency. For instance, under 60\% load and a mixture of both inter- and intra-DC workloads, \tool\ improves the $99^{th}$ percentile FCT by 31\% and 30\% compared to BBR+MPRDMA and Gemini, respectively.

\section{Coexistence among inter- and intra-DC traffic: challenges and opportunities}
\label{sec:motiv}

This section outlines the challenges introduced by the inherent heterogeneity between intra-DC and inter-DC propagation delays.




\subsection{Diverse congestion feedback granularity}
\label{delay_feed}
Typically, the workload inside a datacenter is comprised of both intra-DC and inter-DC traffic. Meanwhile, there exists a huge gap between the congestion feedback granularity of inter- and intra-DC flows due to the significant difference between their propagation delays \cite{gemini}. For instance, considering intra-DC RTT of 10$\mu$s \cite{timely, swift} and inter-DC RTT of 10ms \cite{gemini, anulus}, every inter-DC RTT corresponds to 1000 intra-DC RTTs. This means that intra-DC flows can \textit{potentially} receive congestion signals, \eg ECN, $1000\times$ more frequently than inter-DC flows.
Therefore, in cases of network congestion caused by a mixture of inter- and intra-DC traffic, the intra-DC flows adjust their rates more frequently than inter-DC flows which can potentially victimize intra-DC traffic and hurt flow-level fairness.

Furthermore, as shown in Figure \ref{fig:phantom_advg} \encircle{A} in \S\ref{sec:design_goals}, the large gap between the propagation delay of inter- and intra-DC traffic can cause network under-utilization and queue occupancy oscillations in steady state. Specifically, the congestion feedback for inter-DC flows can be piggybacked to the senders long after the actual congestion had been resolved by intra-DC flows adjusting their rates. In such scenario, reducing the send rate of inter-DC flows upon receiving the congestion signals creates long periods of under-utilization before the flows ramp up and fill the excess network capacity.


\subsection{Heterogeneous hot spots} \label{congestion_signal_intro}
Commodity switches, especially those deployed in inter-datacenter WAN, limit the signals used for congestion detection \cite{gemini}. Therefore, packet loss, delay, and Explicit Congestion Notification (ECN) are commonly used to detect congestion both inside and between datacenters \cite{swift, timely, dcqcn, d2tcp, dctcp, gemini, mprdma}. Solely relying on packet loss and packet re-transmission for detecting congestion and reacting to it imposes significant extra latency as it is only triggered when the switch buffers are extremely congested \cite{swift, dctcp, d2tcp}. 

With ECN as the congestion signal, it is difficult to properly set the ECN marking threshold when having mixed inter-DC and intra-DC traffic \cite{gemini}. This is because the buffer capacity of WAN switches can be larger than that of the switches deployed inside datacenters, as is the case in Gemini \cite{gemini}. Additionally, inter-DC Bandwidth Delay Products (BDPs) are usually much larger than intra-DC BDPs. Therefore, inter-DC traffic requires much larger ECN marking thresholds compared to traffic staying within the datacenter. With delay as the congestion signal, it is challenging to distinguish inter-datacenter hot spots from intra-datacenter. Specifically, it is non-trivial to know whether the increased delay indicates extreme congestion in shallow-buffered intra-DC switches or minor congestion in deep-buffered inter-DC switches. Using shallow buffers everywhere could help the delay signal but it would still be a noisy signal due to the large inter-DC latency. Annulus \cite{anulus} introduces Quantized Congestion Notification (QCN) \cite{qcn1, qcn2} to detect early congestion for inter-DC flows. However, it only helps if the congestion happens near source before crossing the datacenter boundary since it relies on sending an early warning on the reverse path from the congestion hot spot to the source.



\subsection{BDP heterogeneity} \label{phantom_intro}

Most reactive congestion control protocols assume a minimum amount of available buffer capacity for proper operation, \eg DCTCP \cite{dctcp} requires the buffer space to be at least 17\% of BDP.
If we consider the latest switches from Broadcom such as the Trident4 \cite{bcm_t4}, we are given $\sim 4$ MB of buffering per port. Assuming 10 ms inter-DC round-trip delay and 400 Gbps link bandwidth, DCTCP requires at least $\sim 100$ MB of buffering per port, which is much larger than today's switching fabric. Accordingly, intra-DC switches cannot support inter-DC traffic that co-exists with intra-DC traffic inside datacenters.
This gap is likely to increase even further as the network bandwidth keeps increasing at a faster rate than the buffering sizes \cite{anulus}. 

Furthermore, having a buffer significantly smaller than the BDP makes it harder to properly assess the extent of the congestion, as small changes in the sending rate can easily result in either over- or under-utilization of the network. Therefore, in \S\ref{sec:design}, we re-purpose phantom queues \cite{phantomQ}, \ie virtual queues that were originally designed to provide low-latency within a datacenter, for inter-datacenter communication to easily match the inter-DC BDP.


\subsection{Inefficient loss handling} \label{failures_erasure_intro}

As discussed previously, most of the message sizes that cross WAN links are bounded by the propagation delay. This has a very important implication: a single packet loss could significantly increase message delivery time since detecting an inter-DC packet loss and retransmitting it is proportional to the large inter-datacenter RTT. To shed more light on the importance of efficiently handling loss for inter-DC flows, we measure the failure rates between pairs of cloud VMs located in different regions of North America. More specifically, we set up a simple RDMA API that sends 320 million 2KiB packets between the pairs of datacenters. The first selected pair of datacenters (Setup 1) has an RTT of $\approx$ 65 \unit{ms} and an average loss rate of $5.01 \times 10^{-5}$ while the second pair (Setup 2) observes an RTT of $\approx$ 33 \unit{ms} and an average loss rate of $1.22 \times 10^{-5}$. We observe that while losses are rare, they can impose significant extra latency. Thus, fast loss resolution is crucial for inter-DC traffic.
In addition to measuring the overall loss rate, we grouped the packets into consecutive chunks of 10 packets and determined the probability of losing more than one packet within the chunks. The results (Table \ref{tab:packet-loss-comparison}) uncovered that link-correlated drops within a chunk exist, implying that a multi-link failure resilient scheme is preferred.


\begin{table}[htbp]
  \centering
    \scalebox{0.9}{
        \begin{tabular}{ccc|cc}
          \toprule
          \multirow{2}{*}{\shortstack{\textbf{Losses Within}\\\textbf{a Block}}} & \multicolumn{2}{c|}{\textbf{Setup 1 (65ms RTT)}} & \multicolumn{2}{c}{\textbf{Setup 2 (33ms RTT)}} \\
          \cmidrule(lr){2-3} \cmidrule(lr){4-5}
           & \textbf{Drops} & \textbf{Loss Rate} & \textbf{Drops} & \textbf{Loss Rate} \\
          \midrule
          1 & 97\,403  & $3.0\times10^{-4}$ & 12\,785 & $4.0\times10^{-5}$ \\
          2 & 23\,984  & $7.5\times10^{-5}$ & 7\,262  & $2.3\times10^{-5}$ \\
          3 & 5\,007   & $1.6\times10^{-5}$ & 1\,560  & $4.9\times10^{-6}$ \\
          \bottomrule
        \end{tabular}%
  }
  \caption{\small{Packet loss information for two datacenter configurations.}}
  \label{tab:packet-loss-comparison}
  \vspace{-4mm}
\end{table}


\section{\tool\ Design Goals}
\label{sec:design_goals}
Before outlining the details of our design in \S\ref{sec:design}, this section highlights the goals we aim to achieve by proposing \tool.

\subsection{Unified congestion control logic}

As discussed in the previous section, in a congested network, there is a massive gap in the granularity at which transports such as Gemini \cite{gemini} react to congestion signals for inter- and intra-DC flows, which can potentially victimize intra-DC flows. At the same time, the scale of AI training tasks is exceeding the resources available in one datacenter, meaning that messages of the same importance can flow both within and across datacenters. Therefore, ensuring bandwidth fairness among inter- and intra-DC flows and fast convergence to the bandwidth fair share is critical \cite{gemini}\footnote{One approach to flow-level fairness is using multiple priority queues, but inter-DC switches may not support them \cite{anulus}. It also fails to address the lack of inter-DC buffer space \cite{anulus} and requires constant tracking of competing flows to apply weighted round-robin scheduling between inter- and intra-DC traffic.}.

\begin{figure}[t]
	\centering
    \includegraphics[width=\linewidth]{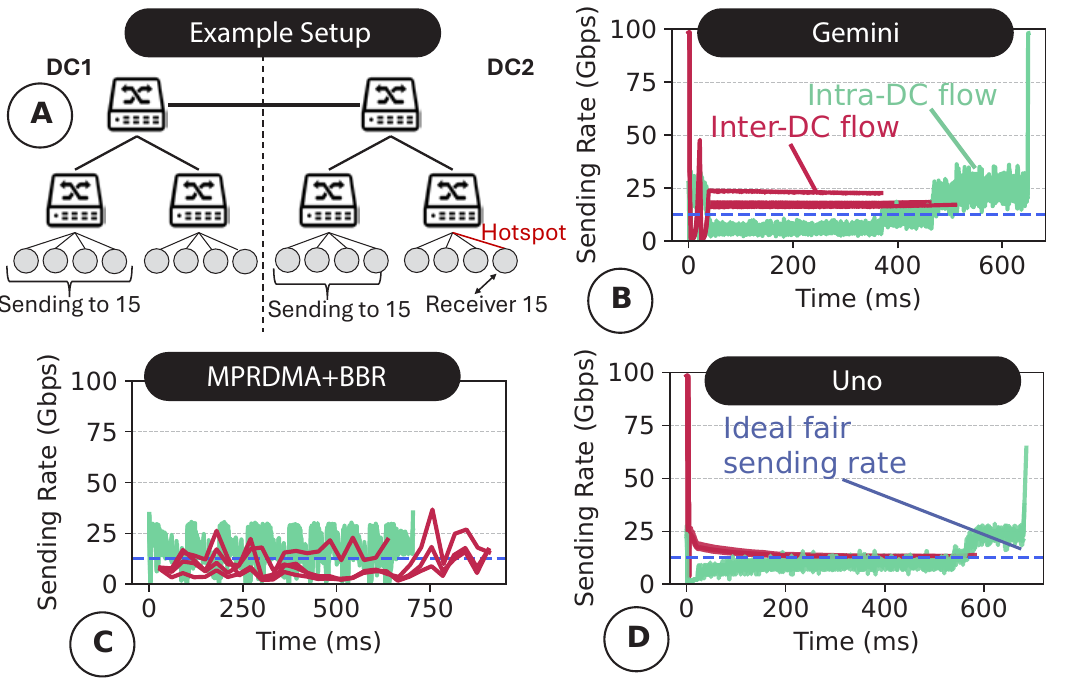}
    \vspace{-6mm}
    \caption{\small{Gemini and MPRDMA+BBR fall short in efficiently achieving bandwidth fairness while \tool\ provides fast convergence to fairness during a mixed incast scenario.}}
    \vspace{-2mm}
    \label{fig:motiv}
\end{figure}

Alas, existing solutions \cite{bbr, anulus, gemini} either do not achieve bandwidth fairness as they separate congestion control for inter- and intra-DC flows, \eg applying BBR \cite{bbr} for inter-DC and DCTCP \cite{dctcp} for intra-DC traffic, or experience slow convergence time because they react to inter- and intra-DC congestion at different granularities. To illustrate this, we simulate two 8-ary fat-tree datacenters \cite{fattree} connected by eight 100 Gbps links via two border switches \cite{gemini}, with inter-DC RTT set 128× larger than intra-DC RTT \cite{gemini}. We create incast by generating four intra-DC and four inter-DC 1 GiB flows toward the same destination and record sending rates for fairness. Figure \ref{fig:motiv} \encircle{A} shows a simplified model of this setup.

We start by measuring rates as we use Gemini \cite{gemini} as congestion control, \ie Figure \ref{fig:motiv} \encircle{B}. While Gemini guarantees convergence to fairness \cite{gemini}, we observe that the convergence occurs so slowly that it outlives the flows' completion times. We repeat the same experiment with BBR's \cite{bbr} and MPRDMA's \cite{mprdma} control loop for inter- and intra-DC flows, respectively. Figure \ref{fig:motiv} \encircle{C} highlights the unfairness among send rates of distinct flows as they are controlled by separate congestion control mechanisms. To address this, in \tool's design, we deploy a unified control loop for both inter- and intra-DC traffic that guarantees fairness while reacting to congestion signals at the same granularity for inter- and intra-DC workload. Our results, \encircle{D}, show that \tool\ converges to fairness considerably faster than Gemini.
To also ensure fast reaction to congestion, \tool\ deploys
\textit{Quick Adapt}, \ie under extreme network congestion, indicated by a sharp drop in the number of ACKed bytes, \tool\ dramatically reduces the send rates to quickly resolve over-utilization.
In \S\ref{sec:evaluation_lab}, we show that \tool\ improves the overall latency against different baselines and under distinct scenarios.




\subsection{Near-zero queuing}

To ensure low latency, especially for small messages, without significantly under-utilizing the network, keeping switch buffers lightly-occupied is essential \cite{dctcp}. However, with empty queues, there is a serious chance of network under-utilization. To avoid this, ECN-based protocols typically keep some packets in the queue with small queue occupancy fluctuations around the ECN marking threshold. However, doing so is challenging with inter-DC traffic in the picture as inter-DC flows can easily overwhelm small commodity switches due to their large BDPs.
To address this, we use phantom queues \cite{lgq, phantomQ}, \ie virtual queues with arbitrary sizes and drain rates that mimic physical queues. They increase occupancy on ingress and drain at a constant rate, typically slightly below the line rate. Intuitively, phantom queues offer two advantages: early congestion signaling (due to lower drain rates) and burst smoothing.

\begin{figure}[htbp]
    \centering
    \includegraphics[width=\linewidth]{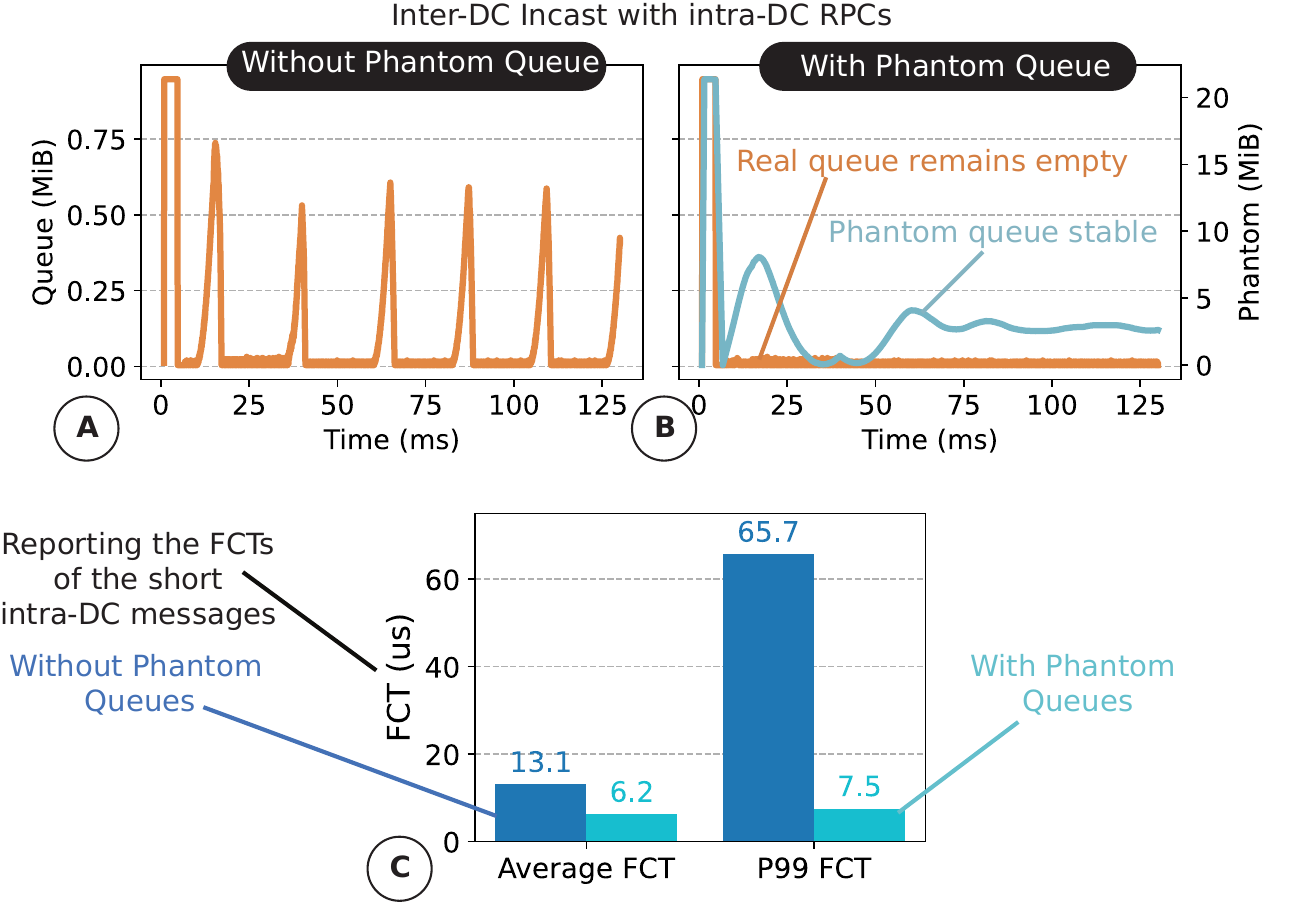}
    \caption{\small{Showcasing the effect of phantom queues for intra-DC traffic. \protect\encircle{A} shows the queue at the incast receiver over time without phantom queues and with phantom queue (\protect\encircle{B}). \protect\encircle{C} shows the FCTs for the intra-DC flows.}}
    \label{fig:phantom_advg}
    \vspace{-3mm}
\end{figure}

To highlight the potential advantages of phantom queues, we simulate a simple scenario where we  initiate long-lived flows from 8 different senders in a local datacenter toward a single receiver in a remote datacenter, creating an incast scenario. In the receiver's datacenter, we also simulate sending several small messages from the "Google RPC" CDF distribution \cite{homa}. Figure \ref{fig:phantom_advg} illustrates the queue occupancies over time with and without phantom queue at the receiver bottleneck and the flow completion times of the small RPC messages. As expected, phantom queues facilitate near-zero queuing that results in $2\times$ and $8\times$ improvement in the average and $99^{th}$ percentile FCT of the RPC messages, respectively.

\subsection{Reliability and Load Balancing}
To mitigate the delay penalties induced by packet loss and retransmissions, explained in \S\ref{failures_erasure_intro}, we adopt Maximum Distance Separable (MDS) \cite{aguilera2005erasure} erasure coding as a proactive countermeasure. In our scheme, data is organized into distinct blocks, each composed of both the original data packets and additional parity packets computed via MDS coding. This “block” represents the minimal unit of encoded data, ensuring that the original information can be fully recovered as long as a sufficient number of packets are received, even if some fail during transit. Given that our experiments reveal that packet losses are not purely random but tend to occur in correlated clusters, the redundancy introduced by MDS coding is crucial. It allows the system to tolerate minor burst losses without waiting for slow retransmission timeouts, thereby maintaining low latency across WAN links. This approach reduces recovery delays and optimizes resource utilization by minimizing unnecessary retransmissions.

While erasure coding solves the problem with certain failure modes, it cannot completely help in all cases. For instance, if we use ECMP routing, \eg with Gemini \cite{gemini} and Annulus \cite{anulus}, if a link goes down, temporarily or permanently, all packets in a block would be lost until the routing table gets updated making it harder to reconstruct the message. To resolve this, we develop our custom load balancing scheme to mitigate the classical shortcomings of ECMP, \eg hash collisions \cite{drill}, while also improving the resilience of erasure coding. We describe the details in \S\ref{lb_explanation}.


\section{\tool: a unified system for both intra- and inter-DC communication}
\label{sec:design}
\begin{figure}[t]
    \centering
    \includegraphics[width=0.36\textwidth]{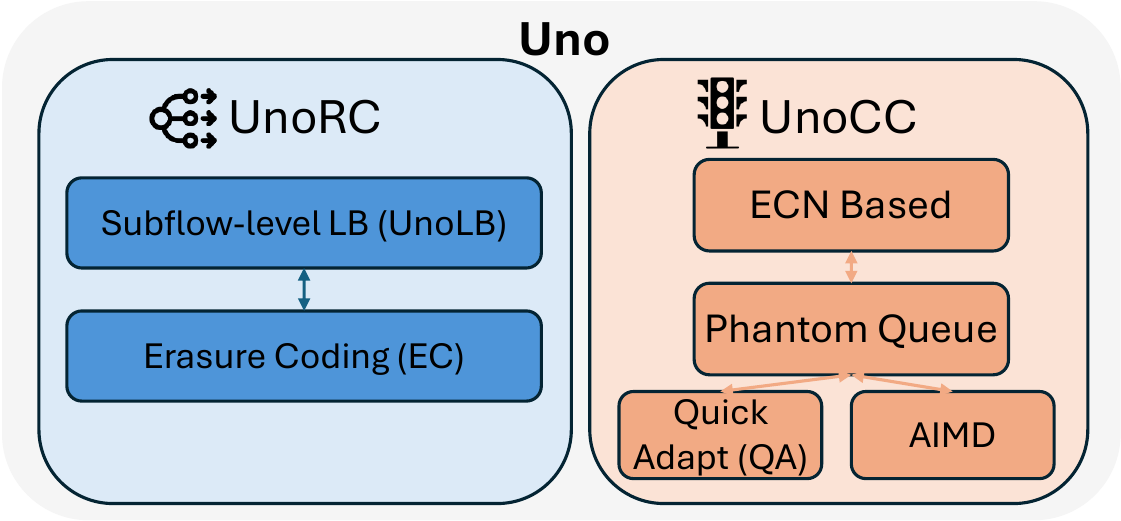}
    \vspace{-4mm}
    \caption{\small{\tool's overall architecture}.}
    \vspace{-4mm}
    \label{fig:architecture}
\end{figure}

Given the goals outlined in the previous section, we design \tool, a unified system that facilitates low-latency and fair communication in both intra-DC and inter-DC environments.
As shown in Figure \ref{fig:architecture}, \tool\ has two components: 1) Congestion Control component (\toolCC) and 2) Reliable Connectivity component (\toolLB).

\textbf{\toolCC} is a window-based congestion control scheme for both intra- and inter-DC traffic that employs Additive Increase Multiplicative Decrease (AIMD) window adjustment to ensure fair bandwidth sharing. To quickly converge to bandwidth fairness, \toolCC\ reacts to congestion signals at the same granularity for intra- and inter-DC flows. It further employs \textit{Quick Adapt}\footnote{Quick Adapt has been previously proposed for communication inside datacenters \cite{bonato2024smarttreps}. However, we tailor it for both intra- and inter-DC communication in this paper.}, which, under extreme congestion (\ie sharp drop in ACKed bytes), promptly reduces the congestion window to quickly alleviate congestion and avoid persistent over-utilization.
To efficiently handle ECN marking in both inter- and intra-DC switch buffers, \toolCC\ is augmented with phantom queues \cite{phantomQ}, \ie
virtual queues with arbitrary sizes and drain rates that mimic physical ones. Delay is used to distinguish physical from phantom queue congestion.

\textbf{\toolLB} combines subflow-level load balancing (UnoLB) with \textit{erasure coding} \cite{erasure, aguilera2005erasure} for inter-DC flows to improve routing performance between datacenters and ensure loss resiliency. The key idea is to use erasure coding to maximize the chances of latency-bound messages getting delivered correctly to the receiver. To do so, we send a certain number of parity packets for every \textit{block}. However, to further improve loss resiliency for a block, we also spread packets of a single block across different paths to maximize the probabilities of successful deliveries even in case of link failures. Finally, we adaptively remove paths from our routing options when we identify them as failed or congested paths which is detected either via a sender-based timeout or a NACK from the receiver.

\subsection{\toolCC}
\label{subsec:designCC}


\begin{figure}[t]
    \centering
    \includegraphics[width=0.44\textwidth]{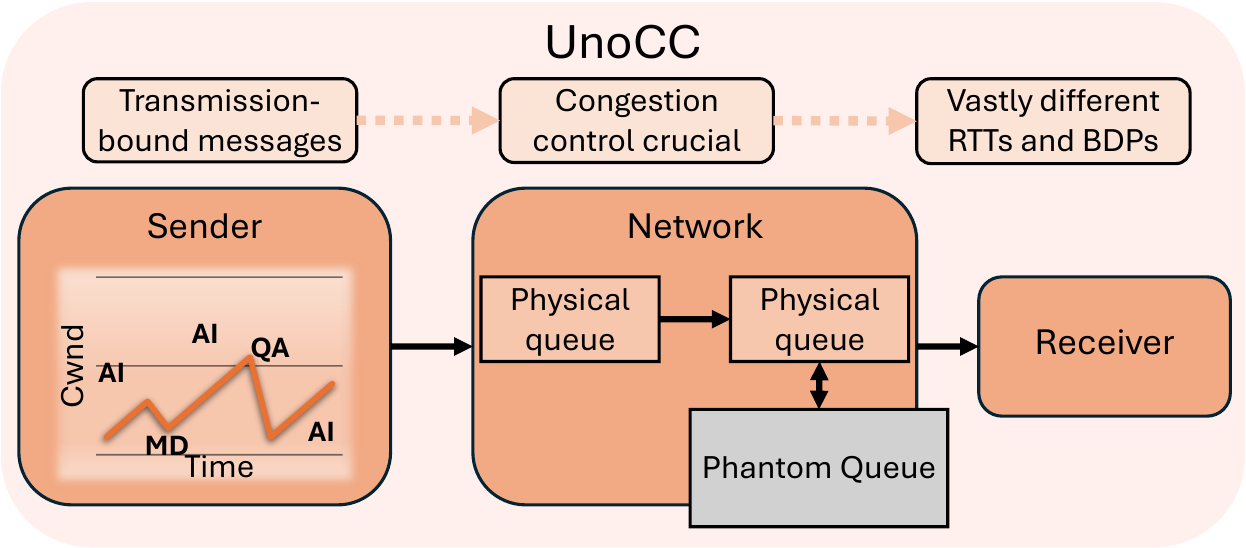}
    \caption{\small{\toolCC's design.}}
    \label{fig:cc}
\end{figure}

As presented in Figure \ref{fig:cc}, \toolCC\ assumes three congestion states for the network: 1) Uncongested, 2) Congested, and 3) Extremely congested. To cope with states 1 and 2, \toolCC\ employs an \textit{AIMD} rate control mechanism that uses ECN
as the congestion signal. \toolCC\ also uses relative delay, \ie $RTT-RTT_{base}$, but only to differentiate between phantom and physical queue congestion events ($RTT$ and $RTT_{base}$ are a packet's measured RTT and the minimum RTT in an uncongested network).
Additionally, \toolCC\ deploys Quick Adapt to facilitate fast reaction to extreme network congestion (State 3).
Algorithm \ref{alg:designCC} outlines distinct mechanisms in \toolCC's design.

\setlength{\textfloatsep}{0.1cm}
\setlength{\floatsep}{0.1cm}



\begin{algorithm}[htpb]
    \footnotesize
    \caption{\small{\toolCC's control loop}}
    \begin{algorithmic}[1]
    \Procedure{OnAck}{}
        \If{$ECN\ not\ marked$} \Comment{Uncongested network (AI)}
            \State $cwnd = cwnd + \alpha \times \frac{bytes\_acked}{cwnd}$
        \EndIf
    \EndProcedure
    \State
    \Procedure{OnEpoch}{}
        \If{$ecn\_fraction > 0$} \Comment{Congested network (MD)}
            \If{$delay == 0$} \Comment{Congestion in phantom queues}
                \State $MD_{scale} = MD_{scale} \times 0.3$ \Comment{Gentle Reduction}
            \Else \Comment{Congestion in physical queues}
                \State $MD_{scale} = 1$
            \EndIf
            \State $cwnd = cwnd \times (1 - MD_{ECN} \times MD_{scale}) $
        \EndIf
    \EndProcedure
    \State
    \Procedure{OnQA}{}
        \If{$bytes\_acked\_in\_qa < cwnd \times \beta$} \Comment{Very congested network (QA)}
            \State $cwnd = bytes\_acked\_in\_qa$\
        \EndIf
    \EndProcedure
\end{algorithmic}
\label{alg:designCC}
\end{algorithm}
\setlength{\textfloatsep}{0.1cm}
\setlength{\floatsep}{0.1cm}

\subsubsection{Additive Increase -- Multiplicative Decrease (AIMD)} 
When an ACK packet arrives and it is not ECN marked,
\toolCC\ increases the congestion window ($cwnd$) by $\alpha \times \frac{bytes\_acked}{cwnd}$. $\alpha$ is the AI factor and is set as a fraction of BDP, \eg $0.001\times BDP$ for our simulations. Thus, after one RTT in an uncongested network, $cwnd$ increases by $\alpha$. Note that $\alpha$ should be scaled depending on the queue size and the degree of incast that a network can support without losses in the steady state.
Unlike AI, that is applied per ACK, MD is applied at most once per \textit{epoch}. Specifically, upon receiving the first ACK of the flow, \toolCC\ stores an epoch activation time ($\mathcal{T}_{epoch}$) for that flow which is initialized to the time of the ACK arrival. Additionally, when a packet is being sent/re-sent, \toolCC\ stores its send/re-send time ($\mathcal{T}_{pkt}$). An epoch terminates when we receive an ACK for a data packet whose $\mathcal{T}_{pkt}$ is $\geq \mathcal{T}_{epoch}$. Upon epoch termination, \toolCC\ increases $\mathcal{T}_{epoch}$ by $epoch\_period$, \ie a time span proportional to the packet's RTT, and re-activates the epoch. Using this approach, we ensure that enough packets are received before deciding to apply MD which gives us better grasp of the network's condition.
\toolCC\ considers an epoch period as congested if any packets has been ECN-marked during the epoch.
When applying MD, \toolCC\ computes the MD factor (\ie $MD_{ECN}$) as $\mathcal{E}\times (\frac{4\times K}{K + BDP})$.
$\mathcal{E}$ represents the exponential weighted moving average (EWMA) of the fraction of ECN-marked packets across epochs, and $K$ is a user-set constant that indicates the extent of \toolCC's reaction to congested epochs.

We select \toolCC's AI and MD factors (\ie $\alpha$ and $MD_{ECN}$, respectively) similar to Gemini \cite{gemini} to achieve guaranteed convergence to fairness.
However, Gemini experiences slow convergence time due to reacting to congestion signals at different granularities for inter- and intra-DC traffic (\S\ref{sec:motiv}). Through extensive empirical experimentation, we found that by reacting to congestion signals at the same granularity for both inter- and intra-DC flows, we can better capture congestion events inside and across datacenters and, thus, considerably improve the speed of convergence to the fair bandwidth share. To this end, we use the same $epoch\_period$, set based on intra-DC RTT, for both inter- and intra-DC flows.
In \S\ref{sec:evaluation_lab}, we quantitatively show that, despite having similar AI and MD factors, \toolCC\ achieves much faster convergence to fairness than Gemini \cite{gemini}.

Finally, since phantom queues have slower drain rates than physical queues, they can signal the sender to slow down more than needed and for too long. To avoid this, if \toolCC\ detects that the physical queues are empty and phantom queues are congested, \ie packets are ECN marked but packet delays indicate no congestion, it employs a gentler reduction (${MD_{scale}}$) by scaling down ${MD_{ECN}}$.

\subsubsection{Quick Adapt (QA)}
Events such as the arrival of new flows or incast can potentially create extreme network congestion. Solely relying on MD for resolving such congestion events is slow and can significantly hurt latency. To avoid this, \toolCC\ deploys the \textit{QA} mechanism. Specifically, once every RTT, \toolCC\ evaluates if the network is extremely congested by checking if
the number of ACKed bytes is considerably low, \ie less than $cwnd \times \beta$ ($\beta$ is the user-set QA ratio).
If the network is deemed as extremely congested, the $cwnd$ is sharply decreased to the number of bytes ACKed during the QA period to quickly match the network's instantaneous capacity. To avoid over-reacting to congestion, after triggering QA, \toolCC\ skips one RTT without triggering any QAs or MDs.


\subsubsection{Phantom Queues}
As discussed in \S\ref{sec:motiv}, efficiently setting ECN marking thresholds for a mixture of inter- and intra-DC traffic is challenging as intra- and inter-DC switch buffer capacities differ, and using only shallow buffers everywhere can also lead to oscillations for modern CCs due to the large inter-DC BDPs. To address this, we use phantom queues \cite{phantomQ}
in conjunction with \toolCC. The phantom queue occupancy increases every time a new packet is enqueued in the physical queue and decreases at a constant rate (\ie draining rate), which is a fraction of the link bandwidth. Note that a phantom queue is easily implementable using a counter that keeps track of its occupancy.

Using phantom queues enables us to correctly mark packets with ECN signals regardless of the physical queue's capacity. As illustrated in \S\ref{sec:motiv}, by properly setting the draining rate of the phantom queues, we practically experience zero queuing at the physical queues as long as the network is at its steady state. Specifically, in steady state, as the phantom queues are drained at a lower speed than the physical queues, the physical queues become empty before the phantom queue occupancy reaches zero.
This is important as it gives extra bandwidth headroom, especially for small and latency-sensitive intra-DC flows. Using phantom queues can potentially penalize large throughput-intensive flows. However, our experiment results show that by setting the phantom queue's draining rate slightly lower than physical queues, \eg 10\% lower, we avoid victimizing large flows while keeping the benefiting small flows\footnote{This result aligns with prior work on phantom queues \cite{phantomQ}.}.

\subsection{\toolLB}
\label{lb_explanation}

\begin{figure}[htbp]
    \centering
\includegraphics[width=0.44\textwidth]{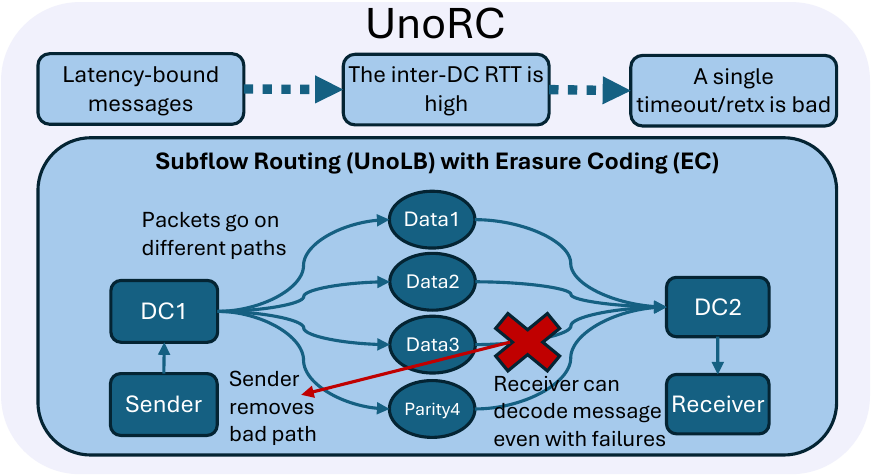}
    \vspace{-2.4mm}
    \caption{\small{\toolLB's design.}}
    \label{fig:lb}
\end{figure}

As shown in Figure \ref{fig:lb}, \toolLB\ has two components: an erasure coding component to enhance reliability, especially under failure, and a simple but effective sub-flow level load balancer (UnoLB).

\textbf{Erasure Coding (EC):} For reliability, each inter-DC message is divided into \textit{blocks} of $n$ packets, with $x$ data and $y$ parity packets. A block can be reconstructed if at most $y$ out of $n$ packets are lost. Upon receiving the first packet of a block, the receiver starts a timer set to the estimated maximum queuing and transmission delay. If the timer expires before enough packets arrive, a NACK is sent to the sender requesting retransmission of the missing block. \toolLB\ applies erasure coding only to inter-DC traffic due to long recovery delays. While EC adds fixed overhead (e.g., 20\%), it reduces packet loss and improves completion times under failure and congestion events, which is crucial in latency-bound scenarios (\S\ref{sec:evaluation_lab}).

\setlength{\textfloatsep}{0.1cm}
\setlength{\floatsep}{0.1cm}



\begin{algorithm}[htb]
\footnotesize
\caption{Pseudocode for UnoLB}\label{algo:unolba}
\begin{algorithmic}[1]

\Procedure{onSend}{\textit{packet}}
    \State $packet[\textit{header.source\_port}] \gets subflow[\textit{index}]$
    \State $\textit{index} \gets (\textit{index} + 1) \bmod \textit{total\_subflows}$
\EndProcedure

\vspace{2pt}
\Procedure{onNackOrTimeout}{\textit{packet}}
    \If{$(now() - last\_reroute) > base\_rtt$}
        \State \textit{update\_subflow}(\textit{packet})
        \State $last\_reroute \gets now()$
    \EndIf
\EndProcedure
\end{algorithmic}
\end{algorithm}
\setlength{\textfloatsep}{0.1cm}
\setlength{\floatsep}{0.1cm}

\textbf{Load Balancing (UnoLB):} Our load balancing scheme uses $n$ subflows and each subflow gets assigned its own path (either via source-based assignment or by changing the source port value for ECMP hashing). By itself, this simple action (somewhat similar to MPTCP \cite{mptcp}) vastly improves the performance due to a decrease in hash collisions. To integrate UnoLB with the reliability aspect (\ie erasure coding), we spread the packets of a block across $n$ subflows. Doing so increases the resilience to link failures. 
Moreover, \toolLB\ switches from \textit{bad paths} when it detects extreme congestion on them.
In particular, upon receiving a NACK (indicating an unrecoverable block) or when a sender timeout occurs (possibly due to lost NACKs caused by failures or corruption), \toolLB\ re-routes the affected flows by randomly selecting a subflow that has recently received ACKs, thereby reducing the likelihood of switching to another congested or failed path.
Algorithm \ref{algo:unolba} presents \toolLB's logic.


\section{Performance Evaluation} 
\label{sec:evaluation_lab}

We use \textit{htsim}, a packet-level network simulator \cite{ndp}, to evaluate \tool\ across distinct workloads and traffic patterns. Our key findings are summarized below:

\begin{itemize}[leftmargin=2em]
    \item \tool\ improves the average and tail latency compared to the state-of-the-art solutions. Specifically, under 40\% load and a mixture of inter-DC and intra-DC flows, \tool\ improves the $99^{th}$ percentile FCT by $1.4\times$ compared to both MPRDMA+BBR and Gemini.
    \item \toolCC\ provides fast convergence to fairness. Particularly, with \toolCC\ under incast events created from various combinations of intra- and inter-DC flows, all flows quickly converge to their fair bandwidth share.
    \item \toolLB\ (\ie UnoLB + EC) further improves the performance under several failure scenarios by up to $3\times$ compared to \tool\ without erasure coding and $2\times$ and $6\times$ compared to RPS and PLB, respectively.
\end{itemize}

\subsection{Simulation Setup} \label{setup_sim}

\begin{figure*}[!ht]
\centering
\includegraphics[width=1.0\textwidth]{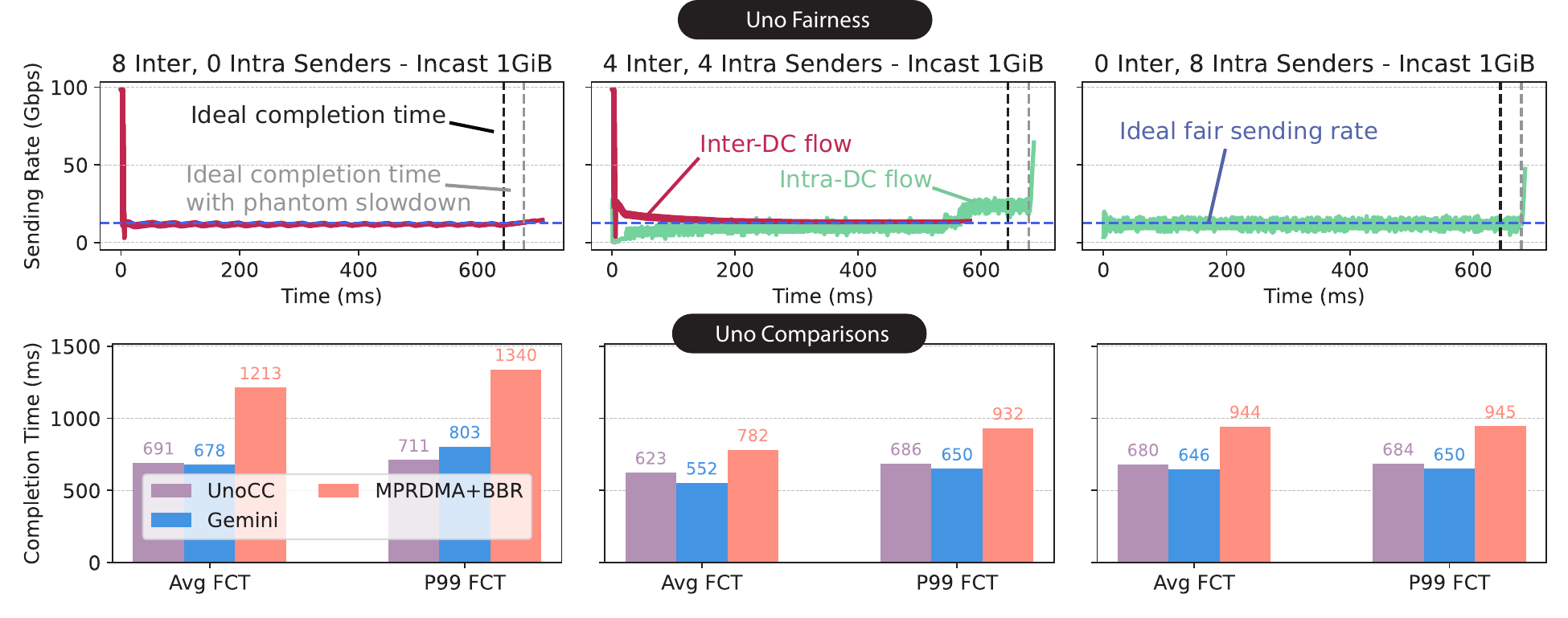}
\vspace{-7mm}
\caption{\small{The top plot showcases the fairness of Uno while the bottom plot shows the performance against other algorithms. The title on top of each plot indicates how many inter- and intra-DC flows take part in the incast.}}
\vspace{-2mm}
\label{fig:incast}
\end{figure*}

\textbf{Topology.} We simulate two 8-ary fat-tree datacenters \cite{fattree}, each consisting of 16 core switches and 8 pods with 4 aggregate and 4 edge switches. Each edge switch is connected to 4 servers. The datacenters are connected through two border switches that are interconnected through eight links. Also, every core switch is connected to a border switch through eight links. Unless stated otherwise, we exploit 100 Gbps links for all our interconnects and set the switch buffer capacities to 1 MiB per port.

\textbf{Microbenchmarks.} We first run microbenchmark experiments similar to those performed in prior works \cite{gemini} to demonstrate the fairness advantages of \tool. Two types of traffic are evaluated: 1) incast traffic originating from different sources and 2) permutation traffic with randomly selected source and destination nodes.

\textbf{Realistic Workload.} The characteristics of intra-DC traffic differ from traffic spanning multiple datacenter networks \cite{flash_pass_alibaba, ratio}. Therefore, we simulate different workloads within and across datacenters. Similar to \cite{gemini}, we use the flow size distributions of Google's web search \cite{dctcp} for generating intra-DC traffic. To generate inter-DC traffic, we exploit the flow size distribution recorded between two datacenters in Alibaba's regional WAN \cite{flash_pass_alibaba}. Unless stated otherwise, the flows' arrival rates are generated based on an exponential distribution, and the rates are scaled to achieve a desired network load. Flow source and destination servers are selected using a uniform random distribution, similar to previous works \cite{dctcp,gemini,flash_pass_alibaba}. The ratio of datacenter to WAN traffic is set to 4:1.

We also use an AI training workload for the inter-DC traffic in one of our experiments. Particularly, we assume a data parallel training strategy \cite{pytorchddp, zerodp} across the two datacenters, where each datacenter has at least one replica of the model being trained. After computing the gradients during the backward pass of each iteration, an Allreduce (or separate Reducescatter and Allgather) collective operation is initiated to synchronize the gradients across the datacenters. Our experiments simulate inter-datacenter training of the Llama 70B model and the parallelization strategy in its technical report \cite{llama3}, which generates periodic traffic bursts of approximately 70-500 MiB per iteration \cite{mixed_prec_training}. The total number of send operations depends on the number of Allreduce groups in the collective.

\textbf{Evaluation metrics.} We measure the mean and tail ($99^{th}$ percentile) Flow Completion Time (FCT) as our main evaluation metric. We also report metrics such as queue occupancy and sending rate.

\textbf{Baseline state-of-the-art (SOTA) approaches.} We compare \tool\ against two major baselines: 1) MPRDMA+BBR (\cite{mprdma}+\cite{bbr}): Widely deployed for enterprise WAN, BBR targets accurate RTT and bandwidth estimation to maximize throughput and minimize latency. Since BBR is only used for inter-DC communication, we combine it with MPRDMA, an ECN-based congestion control scheme, for managing intra-DC communication. 2) Gemini \cite{gemini}: Similar to \toolCC, Gemini is a congestion control protocol designed for both intra-DC and inter-DC communication that exploits ECN and RTT to detect intra-DC and inter-DC congestion, respectively.\footnote{Annulus \cite{anulus}, which works on top of other schemes such as BBR, could also be used to enhance the performance of \tool\ under oversubscribed topologies. However, we leave this add-on for future work.} To evaluate \toolLB, we evaluate it against different routing schemes such as Random Packet Spraying (RPS) \cite{spraying} and PLB \cite{plb}.

\begin{table}[htbp]
  \centering
    \scalebox{0.9}{
        \begin{tabular}{cc}
            \toprule
            \shortstack{\textbf{Parameter}} & \textbf{Default Value} \\
            \midrule
            $\alpha$ (\toolCC's AI factor) & $0.001 \times BDP$ \\
            $\beta$ (\toolCC's QA factor) & 0.5 \\
            K (\toolCC's MD constant) & $\frac{1}{7}\times$intra-DC BDP \\
            Intra-DC RTT & $14 \mu s$ \\
            Inter-DC RTT & $2 ms$ \\
            Phantom queue drain rate & $0.9\times$physical queue drain rate \\
          \bottomrule
        \end{tabular}%
  }
    \caption{\small{Parameter table.}}
    \vspace{-5mm}
    \label{tab:parameter}
\end{table}

\textbf{Parameter settings.} To mark ECN, we use Random Early Detection \cite{red}. Specifically, the packets are never ECN marked as long as the destination queue occupancy is less than the minimum ECN threshold (\ie \textit{MinECNThresh}) and always marked when the occupancy is more than the maximum ECN threshold (\ie \textit{MaxECNThresh}). Otherwise, the probability of marking packets increases linearly. \textit{MinECNThresh} and \textit{MaxECNThresh} are set to 25\% and 75\% of queue capacity. For \tool, we set $\alpha$, $\beta$, and $K$ to 0.1\% of the BDP, 0.5, and $\frac{1}{7}\times$intra-DC BDP, respectively. The phantom queues drain at 90\% of the line rate. Unless stated otherwise, MTU, intra-DC RTT, and inter-DC RTT are set to 4096 B, $14 \mu s$, and $2 ms$, respectively \cite{gemini}. Table \ref{tab:parameter} summarizes the main parameters of our experiments.

\subsection{Simulation results}

\begin{figure}[htb]
    \centering
    \includegraphics[width=\linewidth]{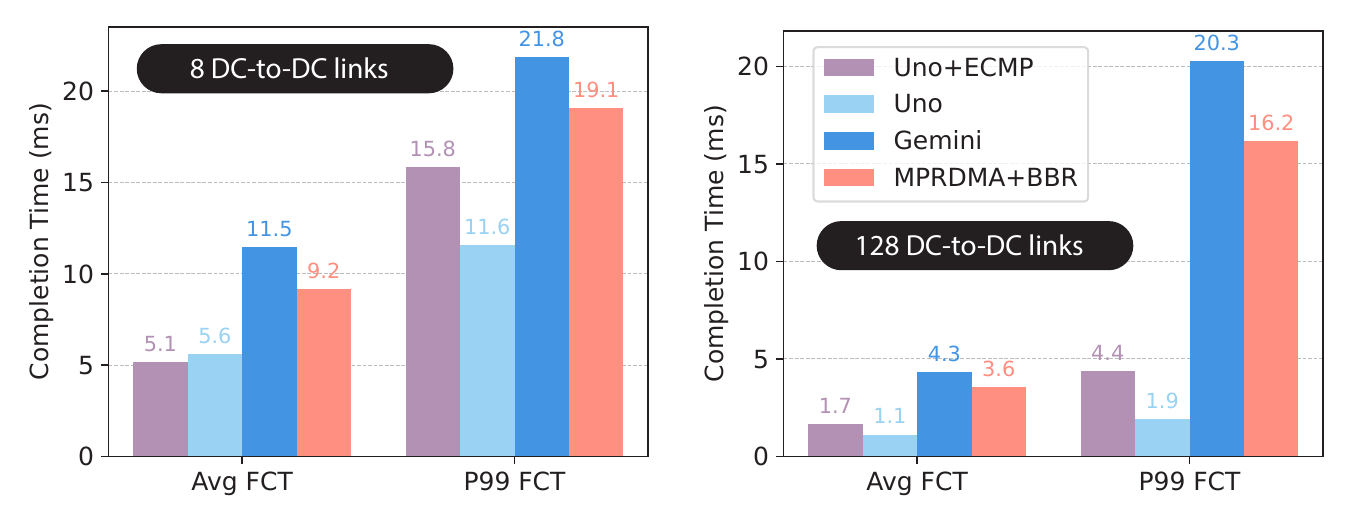}
    \caption{\small{Showcasing results with a permutation workload.}}
    \label{fig:perm}
\end{figure}

\begin{figure*}[!ht]
\centering
\includegraphics[width=1.0\textwidth]{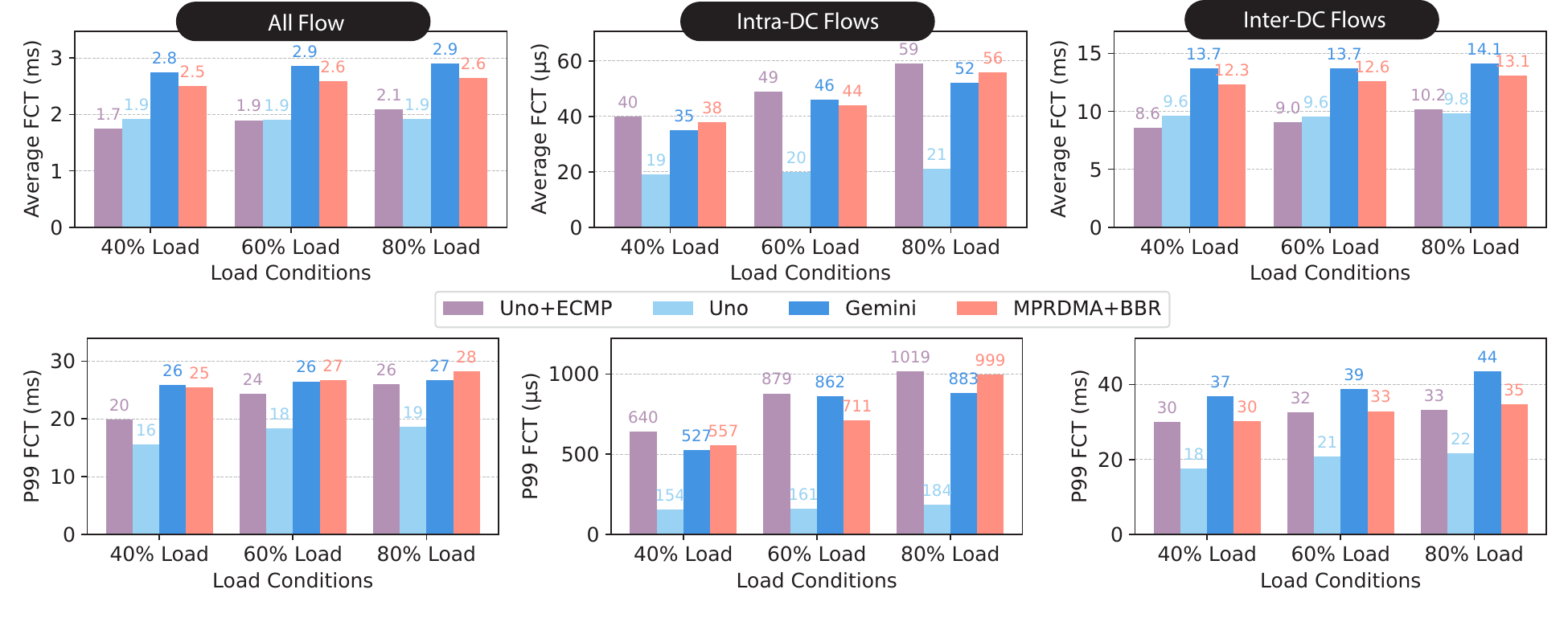}
\vspace{-6mm}
\caption{\small{\tool\ is superior to other baselines under distinct network loads.}}
\vspace{-2mm}
\label{fig:websearch_flashpass_dload}
\end{figure*}


\subsubsection{Micro-benchmarks}
As the first experiment, we simulate \tool\ under intra-DC incast, inter-DC incast, and a mixture of the two. In all scenarios, we generate a total of eight 1 GiB flows. Also, we use packet spraying for all schemes as load balancing has a negligible impact under receiver-side incast. In Figure \ref{fig:incast}, we present \tool's fairness by plotting the sending rate of the flows involved in the incast and evaluate its performance against other baselines. Our findings show that \tool\ outperforms or matches the performance of the other algorithms in all three scenarios and achieves near-ideal latency. Moreover, using \tool, in all scenarios, the send rates of inter- and intra-DC flows quickly converge to their fair bandwidth share, highlighting \tool's fast convergence to fairness.

We then consider a permutation scenario where each sender is sending to another randomly selected node (within the same DC or cross DCs). Since this could result in a lot of inter-DC communication (easily overwhelming the eight inter-DC links), we showcase two scenarios: one where the topology is as-is with 800Gbps of inter-DC bandwidth and another scenario where the inter-DC links are fully provisioned. Here we showcase Uno with ECMP and Uno with our custom load balancing solution (UnoLB) as part of UnoRC. In Figure~\ref{fig:perm}, we show the two results. Under the same ECMP load balancing assumption, Uno is still significantly better than the alternatives in both scenarios. The gap between \tool\ and the other approaches further increases with UnoLB. As expected, average FCTs are higher when fewer links connect the DCs.

\subsubsection{Realistic workloads} \label{realistic_workloads_description}
To test \tool\ against realistic workloads, we generate intra- and inter-DC traffic using Google's web search \cite{dctcp} and Alibaba's WAN \cite{flash_pass_alibaba} traffic distributions, respectively, and compare \tool+ECMP (\ie \toolCC\ deployed beside ECMP for load balancing) and \tool\ (\ie \toolCC+\toolLB) against other baselines.
Figure \ref{fig:websearch_flashpass_dload} presents the results. We observe that, compared to Gemini and MPRDMA+BBR, \toolCC\ reduces both average and tail latency of inter-DC flows but slightly increases the latency of intra-DC web search flows. This is because phantom queues' drain rates are lower than physical queues which can occasionally hurt the latency of intra-DC flows. However, despite marginally increasing the latency of intra-DC flows, \toolCC\ manages to improve the overall latency by providing better congestion management and faster convergence to fair bandwidth share, \eg under 40\% load, \toolCC\ improves mean latency by 30\% and 37\% compared to MPRDMA+BBR and Gemini, respectively. Additionally, by achieving balanced load and loss resiliency on top of \toolCC, \tool\ manages to reduce the latency of both intra- and inter-DC flows compared to other baselines. For example, compared to MPRDMA+BBR and Gemini under 40\% load, \tool\ decreases the tail FCT by $4.4\times$ and $5.3\times$ for intra-DC flows and by  $1.7\times$ and $2.1\times$ for inter-DC flows, respectively.


\begin{figure}[ht]
    \centering
    \includegraphics[width=0.46\textwidth]{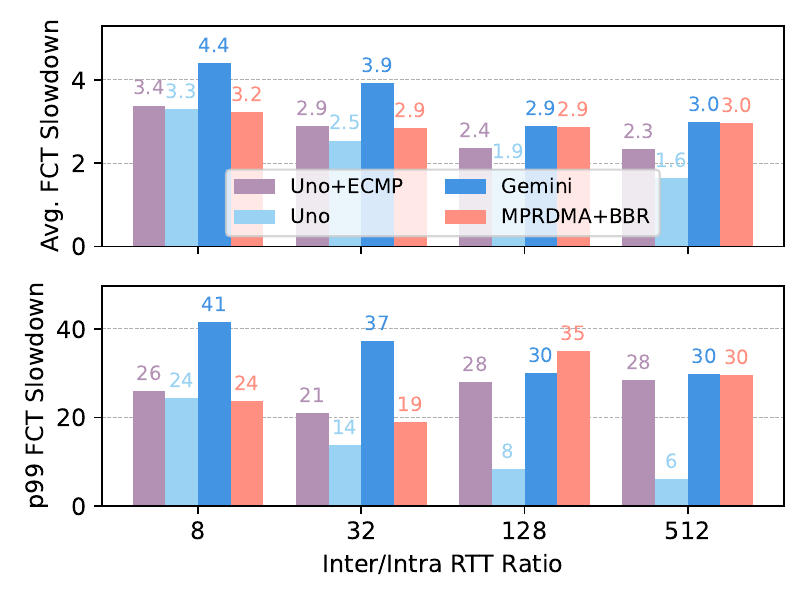}
    \vspace{-2mm}
    \caption{\small{\tool\ improves latency when there exists a significant gap between intra- and inter-DC RTTs.}}
    \label{fig:websearch_flashpass_ddelay}
\end{figure}

We also evaluate \tool\ against different inter-DC propagation delays. To this end, we repeat the previous experiments with 40\% network load and gradually increase the propagation delay of inter-DC links. Figure \ref{fig:websearch_flashpass_ddelay} reports the FCT slowdown ratios of distinct schemes as we increase the ratio of inter-DC minimum RTT to intra-DC minimum RTT from 8 to 512 by increasing the inter-DC propagation delay (intra-DC minimum RTT = $14\mu s$). We observe that, when the gap between intra- and inter-DC RTT is small, \tool\ is slightly outperformed by MPRDMA+BBR due to the phantom queue slowdown. However, as the RTT ratio increases, and gets closer to ratios observed in today's networks (\eg Figure \ref{fig:poster}), \tool\ significantly outperforms MPRDMA+BBR and Gemini. In particular, when the RTT ratio is 512, \tool's tail FCT slowdown is $5\times$ lower than MPRDMA+BBR and Gemini. This shows that the existing solutions fall short in efficiently handling the big gap between intra- and inter-DC delays, while \tool\ is well-designed for it.

\begin{figure}[ht]
    \centering
    \centering
    \includegraphics[width=0.46\textwidth]{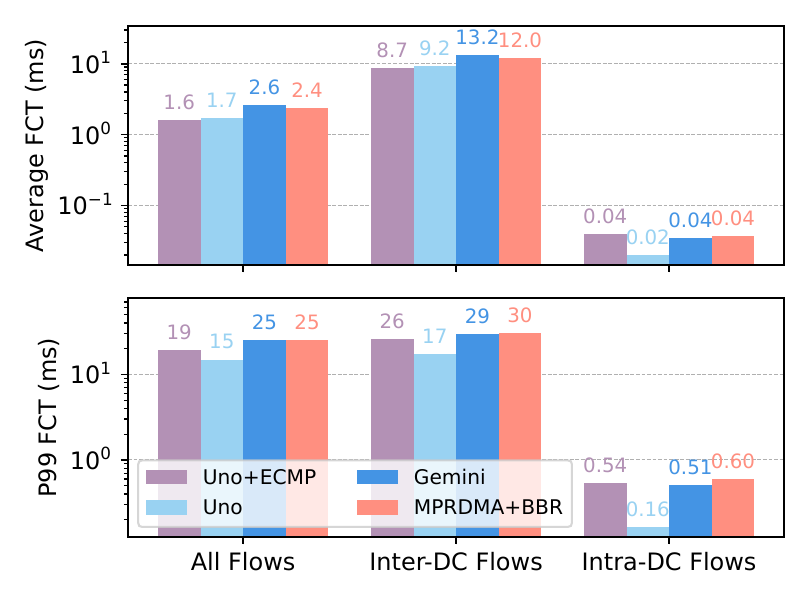}
    \vspace{-2mm}
    \caption{\small{\tool\ is superior to MPRDMA+BBR and Gemini even when the queue capacities inside and across datacenters differ.}}
    \label{fig:different_queue}
\end{figure}

Lastly, we evaluate \tool's performance under realistic workloads and distinct inter- and intra-DC queue sizes. We re-run experiments under 40\% load using shallow- and deep-buffered switches inside and across datacenters, respectively. Particularly, we set intra- and inter-DC queue sizes to $\approx175$ KiB (\ie intra-DC BDP) and $\approx2.2$ MiB (\ie $0.1\times$ inter-DC BDP) per port. Figure~\ref{fig:different_queue} shows average and tail FCTs. Consistent with prior results (\ie Figure~\ref{fig:websearch_flashpass_dload}), \tool+ECMP lowers overall FCT by reducing inter-DC completion times with only slight intra-DC latency increase, while \tool\ improves both: compared to Gemini, tail FCT drops by $3.1\times$ (intra) and $1.7\times$ (inter); versus MPRDMA+BBR, it drops by $3.6\times$ (intra) and $1.8\times$ (inter).

\begin{figure*}[!ht]
\centering
\includegraphics[width=\textwidth]{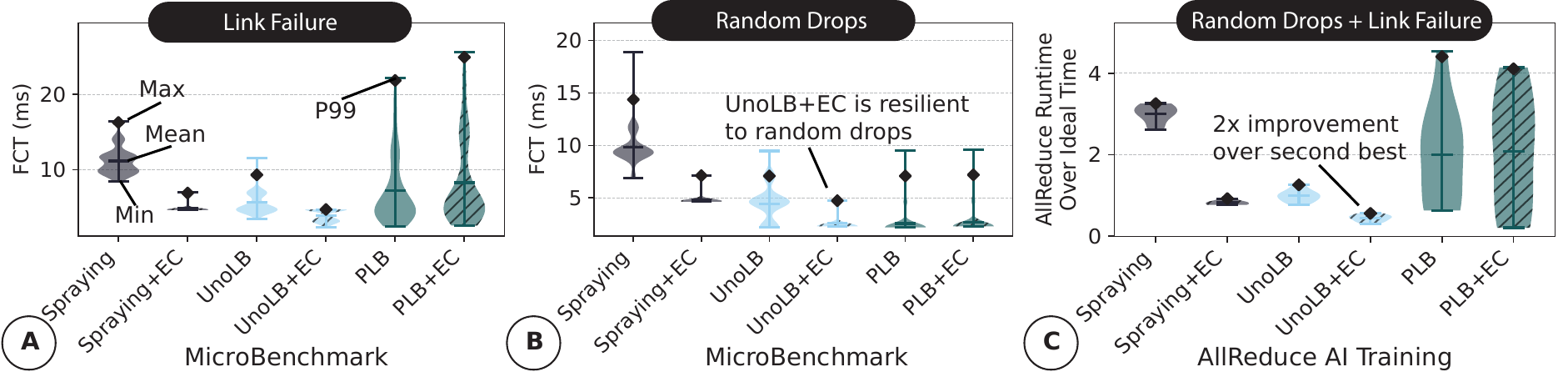}
\caption{\small{Uno's performance under distinct failure scenarios and different workloads.}}
\label{fig:failures}
\end{figure*}

\subsubsection{Failure Scenarios}
We now move to the evaluation of the reliability (RC) aspect of Uno. To do so, we evaluate Uno's performance under different failure scenarios. Since we have already evaluated the effectiveness of UnoCC on its own against other state-of-the-art algorithms, we now focus only on different variants of Uno when it comes to load balancing and reliability and use UnoCC as congestion control for all experiments. In particular, we compare three different state-of-the-art load balancing schemes: packet spraying \cite{spraying}, PLB \cite{plb} and UnoLB. We exclude ECMP since we already evaluated the superiority of multipathing schemes in the previous section and because ECMP is oblivious to network failures.

For erasure coding (EC), we note that the choice of the correct block size and number of parity packets depends on the initial assumptions of the network. Higher failure rates require more redundancy but also introduce more overhead. Although there is no single \textit{one-size-fits-all} answer, we ran several experiments and decided to use a \textit{(8, 2)} scheme, \ie 10 packets in a group with 8 data packets and 2 parity packets.

The first experiment that we run consists of failing one of the eight border links while simulating latency-sensitive 5MiB flows between two datacenters that can theoretically saturate all the inter-DC bandwidth. Since a single run can be heavily influenced by the initial path selection, we re-run every experiment 100 times to reduce biases and use violin plots to show detailed statistics across runs. Figure~\ref{fig:failures} \protect\encircle{A} presents the results. We observe that \tool\ outperforms both spraying and PLB with and without EC. This is due to \tool's ability to adaptively avoid problematic links and intelligently distribute packets within the same block.

Then, we move to a different failure scenario and simulate random loss events based on our real measurements from Table~\ref{tab:packet-loss-comparison}. We implement and simulate the full failure logic comprehensive of the correlation between different failure events. In Figure~\ref{fig:failures} \protect\encircle{B}, we show the results for this scenario considering a single flow across datacenters. In this case, blocks are only dropped in the unlikely event that three or more packets are dropped within a block. What we observe is that Uno is mostly matching packet spraying, as expected, and outperforming PLB both with and without EC. This is because PLB sticks to one path for a given flow, and when a link becomes temporarily flaky, it negatively affects the entire block, resulting in a higher runtime for the worst-case scenario.

Finally, in Figure~\ref{fig:failures} \protect\encircle{C}, we simulate the training strategy described in \S\ref{setup_sim}. In this experiment, we simulate both link failures and random drops. We generate 100 training iterations based on CrossPipe~\cite{crosspipe} and focus on the inter-DC communication. We report the ratio of the measured Allreduce collective runtime per training iteration to the ideal runtime. We observe that Uno consistently outperforms other baselines both with and without EC. Specifically, with EC, \tool\ performs over $2\times$ better than the second best algorithm and only 30\% slower than the ideal completion time that assumes no ECMP collisions or random drops.

\section{Discussion}
\label{sec:discussion}

\textbf{Leveraging modern datacenter congestion control for inter-DC traffic.} Using MPRDMA+BBR, we show the drawbacks of separating intra- and inter-DC control loops. While alternatives like HPCC \cite{hpcc} and PowerTCP \cite{powerTCP} exist, they too suffer from fairness issues due to this separation. Unfortunately, most SOTA intra-DC congestion control protocols, \eg \cite{hpcc, bolt, ndp, dctcp, mprdma, swift}, rely on fast RTT feedback and specialized switch support (\eg INT and packet trimming), making them impractical across inter-DC environments. Additionally, protocols like PowerTCP and HPCC use single-pathing, limiting their ability to balance load. 
In contrast, \tool\ applies congestion control uniformly across intra- and inter-DC traffic, relies on widely supported ECN \cite{dctcp, ecn_easy}, and employs multi-pathing to achieve low latency, high reliability, and fairness.

\textbf{Hardware implementation.} While we showcase Uno’s potential in simulations, we design it with hardware feasibility in mind:

\textit{\toolCC}’s key operations (\ie AIMD and QA) are easily deployable in the Linux kernel, similar to other widely deployed protocols like TCP/DCTCP. For the congestion signal, UnoCC only uses ECN, which is commonly supported by today’s switches \cite{dctcp, ecn_easy}. Phantom queues, already supported by several vendors \cite{ph_impl1, ph_impl2} (such as in Cisco Nexus 5548P \cite{nexus_5548}) and implementable with simple packet-increment/decrement counters, are also easily deployable on most switches. Additionally, \tool\ uses hardware pacing for congestion control, which is commonly available at the sender NIC \cite{anulus, swift, phantomQ}.

\textit{\toolLB} is deployable as software shim layers: the sender inserts parity packets and the receiver decodes once enough packets arrive, using a coarse software timer suited for long-latency links. Finally, \tool's re-routing can be implemented by either changing the IPv6 flow-label \cite{plb} or updating the UDP source port as described in Ultra Ethernet \cite{hoefler2025ultraethernetsdesignprinciples}.

\section{Related Work}
\label{sec:related}
\textbf{Congestion within and across datacenters.}
Most of the existing proposals on congestion control either focus on intra-DC networks or inter-DC WANs and very little research has been done on simultaneously addressing intra- and inter-DC congestion \cite{gemini}. To address intra-DC congestion, existing body of work typically relies on ECN \cite{dctcp, dcqcn, d2tcp, phantomQ}, delay \cite{timely, swift}, or receiver-driven transmission \cite{homa, expresspass, phost, ndp}. WAN congestion management proposals \cite{bbr, tan2006compound, vegas, copa}, on the other hand, typically use delay to detect WAN congestion due to the limited congestion detection support from WAN switches. Gemini \cite{gemini} is among the first proposals that tries to bring inter- and intra-DC congestion control together as a system by using ECN and delay to detect intra- and inter-DC congestion, respectively. However, as presented in \S\ref{sec:evaluation_lab}, Gemini suffers slow convergence to bandwidth fairness. Annulus \cite{anulus} tries to address congestion near source but does not effectively handle congestion that occurs far from the source and closer to the destination. Moreover, Annulus works on top of other protocols, thus not a standalone solution on its own. \tool, however, acts as an effective system for both intra- and inter-DC environments by ensuring fast reaction to congestion, bandwidth fairness, and loss resiliency.
HULL \cite{phantomQ} and LGC-ShQ \cite{lgq} have tried using phantom queues but limited to the datacenter networks.

\textbf{Addressing reliability and load balancing.}
Throughout the years, many protocols have been proposed for load balancing inside datacenters, spanning across different granularity: per flow, per sub-flow, or per packet. Protocols such as ECMP, Hedera \cite{hedera}, MicroTE \cite{microte}, Flowcut \cite{bonato2025flowcutswitchinghighperformanceadaptive} work on per-flow level but they either suffer from hash collision or require complex global controllers. Solutions working on sub-flow level load balancing try to mitigate the above issues. FlowBender \cite{flowbender}, PLB \cite{plb}, CONGA \cite{conga} significantly improve routing performance over ECMP. However, they can still be prone to collisions since they use one given path at a time or require specialized hardware as is the case of CONGA. Finally, packet-level load balancers, such as RPS \cite{spraying}, Drill \cite{drill}, and REPS \cite{bonato2025repsrecycledentropypacket}, provide the best absolute performance but require out-of-order support at the receiver and complicate loss detection. UnoLB tries to get the best of both worlds by using multiple sub-flow at any given time (similar to MPTCP \cite{mptcp}), but also by adaptively re-routing flows away from congested paths. This ensures good load balancing performance while limiting the amount of out-of-order packets.
Besides load balancing, erasure coding also improves loss resiliency. Specifically, erasure coding enhances data reliability by splitting the data into fragments and adding strategic redundancy, allowing recovery even with some failures \cite{erasure, erasure1, erasure2}. In Cloudburst \cite{cloudburst}, the idea of multi-pathing with erasure coding was explored but focusing on intra-DC communication.

\section{Conclusion}
\label{sec:conclusion}
Simultaneously handling congestion both inside and across datacenters is challenging due to the huge gap between intra- and inter-DC propagation delays. To address this, we proposed \tool, a unified system for efficient DC communication with two components: (1) \toolCC, a congestion control scheme ensuring fairness and fast reaction, and (2) \toolLB, a reliable routing mechanism combining subflow-level load balancing and erasure coding. Through extensive simulations, we showed that \tool\ improves latency and fairness compared to the state-of-the-art solutions such as BBR and Gemini. 
Moreover, we illustrated the effectiveness of \tool\ in ensuring loss resiliency when encountering failure and against different load balancers.

\begin{acks}
This project was funded by the NSF CNS NeTS grant (No. 2313164), Sapienza University Grants ADAGIO and D2QNeT (Bando per la ricerca di Ateneo 2023 and 2024), and the European Research Council (ERC) under the European Union’s Horizon 2020 research and innovation program (grant agreement PSAP, No. 101002047). We also thank the Swiss National Supercomputing Center (CSCS) for providing the computational resources used in this work.
The authors used ChatGPT to only assist with editing and quality control throughout this manuscript (ideas and content remain original).
\end{acks} 

\bibliographystyle{ACM-Reference-Format}
\bibliography{references}

\end{document}